# Communities detection in complex network and multilayer network systems: A flow approach

Olexandr Polishchuk

*Pidstryhach Institute for Applied Problems of Mechanics and Mathematics, National Academy of Sciences of Ukraine, Naukova str, 3"b", Lviv, 79060, Ukraine*

**Abstract**
A flow approach to community detection in complex network and multilayer network systems is proposed. Two methods have been developed to search for communities in a network system (NS). The first of them is based on the calculation of flow influence parameters of NS's subsystems, selected according to the principle of nesting hierarchy. The second method uses the concept of flow core of network system. Two methods are also proposed for community detection in multilayer network system (MLNS). The first of them is based on the concept of MLNS aggregate-network and subsequent allocation of its flow core. The second method uses the concept of flow core of the process of intersystem interactions in general. All developed methods are based on the use of flow criterion that the selected group of nodes really forms a community. The results of application of developed approaches are illustrated by examples for which known methods are ineffective.

**Keywords**
Complex network, Network system, Intersystem interactions, Flow model, Hierarcy, Flow core, Aggregate-network, Influence, Community

## 1. Вступ

Однією з важливих проблем, яка досліджується у теорії складних мереж (ТСМ), є пошук груп взаємопов'язаних вузлів, ідентифікація яких сприяє кращому розумінню принципів організації структури та процесів функціонування складних мережевих систем. У реальних МС найбільш поширеними групами є так звані спільноти – підмережі, зв'язки між вузлами яких є щільнішими та сильнішими, ніж між ними та іншими вузлами мережі [1, 2]. Спільноти існують у фізичному світі, живій природі, економіці, на транспорті, у міській інфраструктурі тощо [3, 4]. У людському соціумі спільнотами можна вважати громадські організації, політичні партії, релігійні конфесії, національні діаспори, групи в соціальних мережах [5, 6] і т. ін. Натепер основна увага приділяється розробленню методів пошуку спільнот, які базуються на структурних характеристиках мережевих систем – найменшому розрізі, ієрархічній кластеризації, оцінці модулярності або ентропії, спектральних властивостях мережі чи випадковій ході тощо [7, 8]. Не менш важливою та складною є задача пошуку спільнот у БШМС, які описують процеси міжсистемних взаємодій у надсистемних утвореннях різних типів [9, 10]. У цьому випадку зазвичай також використовуються перераховані вище методи і підходи [11]. Основним недоліком відомих методів пошуку спільнот поряд із обчислювальною складністю та ресурсоємністю є відсутність достовірного теоретично обґрунтованого критерію того, що визначена будь-яким із цих методів група вузлів дійсно утворює спільноту, адже якщо термін «щільність» мережі у ТСМ є достатньо зрозумілим і легко обчислюється за відомими формулами, то поняття «сильніший» або «слабший» зв'язок зі структурного погляду не є достатньо чітким та однозначним [8]. Саме ця обставина вимушує інколи використовувати візульні методи дослідження [12]. Додаткова вада існуючих методів полягає у тому, що вони зазвичай спрямовані на пошук вже сформованих та достатньо стабільних порівняно великих за складом спільнот, але не відстежують появу таких спільнот у мережі та їх швидкий розвиток (збільшення, зменшення, зникнення). Навіть динамічні структурні моделі, тобто моделі, які враховують зміни в структурі МС та БШМС з часом, загалом не в змозі вирішити цю проблему [4]. У той же час, в сучасному суспільстві відбувається чимало важливих та масових подій, організованих спільнотами різного спрямування, перебіг яких обмежується кількома тижнями і навіть днями. Велика кількість існуючих методів пошуку спільнот у МС та БШМС свідчить про неабиякий інтерес до цієї проблематики та її важливість у системних дослідженнях [13]. Мета цієї статті полягає у розробленні на підставі потокових моделей складних мережевих та багатошарових мережевих системах критеріїв та методів пошуку спільнот у таких утвореннях.

## 2. Динаміка формування та розвитку спільнот у людському суспільстві

Одним із найбільш цікавих та актуальних об'єктів дослідження у ТСМ є спільноти, які виникають у людському суспільстві та тим або іншим чином впливають на його розвиток [6, 14]. Починаючи із первісних племен, такі спільноти часто відігравали значну як позитивну, так і негативну роль в історичному процесі. Формування світових релігій, утворення нових держав, зміна суспільно-економічних формацій завжди починалися із невеликих, але сильно вмотивованих і достатньо активних груп однодумців. Зародження та діяльність таких груп зазвичай позитивно впливали на розвиток суспільства (спільноти колекціонерів створювали музеї та бібліотеки, любителів культури – філармонії а картинні галереї, науковців – університети та дослідницькі лабораторії і т. ін.). Водночас, поява та поширення расистських, фашистських, комуністичних та інших людиноненависницьких ідеологій мали негативний вплив на перебіг історичних процесів. Події останніх років свідчать, що загроза повторення подібних явищ, причому із значно катастрофічнішими наслідками, нікуди не зникла. Поряд із цим постійно виникали і виникають різнорідні терористичні (Аль-Каїда, ІДІЛ), хакерські (Anonymous, LockBit), організовані кримінальні (мафія, наркокартелі) угрупування та релігійні секти (Храм Народів, Аум Сінрікьо), які тією або іншою мірою впливали і часто натепер впливають на суспільну безпеку та спокій громадян. Постійно виникають порівняно невеликі спільноти, які породжують суїцидальні настрої у підлітків, змушують їх організовувати одночасні масові бійки у багатьох містах кількох країн світу, розвивають їх зневіру у своєму майбутньому, спокушають на споживання наркотичних речовин або залучають до екстремістських організацій різного роду. Виявлення подібних спільнот має не тільки науковий інтерес, але й велике суспільне значення, оскільки припинення їх діяльності перед переходом до конкретних дій дозволяє уникнути багатьох жертв та поламаних доль.

Поширення та розвиток світових релігій продовжувався століттями, нацистських та комуністичних ідеологій – десятиліттями, різнорідних злочинних угрупувань – роками. В сучасному світі із розвитком інформаційних та комунікаційних технологій утворення спільнот може зайняти дні і навіть лічені години. Тобто, якщо раніше подібні процеси займали роки, десятиліття і навіть століття та підштовхувалися серйозними кризовими ситуаціями, такими як війни, голод, епідемії небезпечних інфекційних захворювань, то натепер з використанням соціальних мереж народження та активація спільнот може здійснюватись дуже швидко. Зазвичай такі процеси провокуються неправильними політичними та економічними рішеннями або вчинками, які збурюють соціальну свідомість (порушення прав людини, неадекватні дії поліції або влади і т. ін.). Лише початок 21 століття переповнений такими подіями – Майдани в Україні у 2004 та 2013 роках, революції в Грузії, Киргизстані, Лівії, Тунісі, політичні збурення в Казахстані, Білорусі та Франції тощо. Визначальною особливістю цих подій була швидкість об'єднання великих груп людей та їх масових виступів, що було практично неможливим у «доінформатизаційні» часи. При цьому, спільноти виникали саме у громадянському суспільстві, а соціальні мережі, як інструмент інформаційно-комунікаційних технологій, служили засобом, який сприяв їх якнайшвидшому формуванню. Однак, цей засіб дає змогу кількісно відстежувати процес народження та розвитку таких спільнот.

Спільноти можуть існувати як в окремих шарах-системах БШМС, так і в процесі взаємодій між ними, постійно виникаючи, поєднуючись, перекриваючись або нівелюючи одна одну. Тому для кращого розуміння процесів міжсистемних взаємодій пошук спільнот необхідно здійснювати як в окремих шарах, так і в БШМС загалом. Спільноти у сучасному світі, зокрема громадянські та соціальні, – це зазвичай достатньо динамічні структури, які можуть як швидко з'являтися і розвиватися, так і швидко зникати, і методи виявлення таких утворень повинні враховувати цю особливість. Структурні моделі складних мережевих систем та міжсистемних взаємодій зазвичай не в змозі вирішити цю проблему. Тому динамічні моделі, які описують процеси функціонування МС та БШМС, стають надзвичайно важливими. Розглянемо потоковий підхід для виявлення спільнот у таких системах та міжсистемних утвореннях.

## 3. Структурна та потокова моделі багатошарової мережевої системи

Структура міжсистемних взаємодій описується багатошаровими мережами (БШМ) та відображається у вигляді [1]

$$G^M = \left( \bigcup_{m=1}^{M} G_m, \bigcup_{m,k=1, m \neq k}^{M} E_{mk} \right), \tag{1}$$

де $G_m = (V_m, E_m)$, $G_m \in R^n$, $n = 2,3$, визначає структуру $m$-го мережевого шару БШМ; $V_m$ – множина вузлів мережі $G_m$; $E_m$ – множина зв'язків мережі $G_m$, $E_{mk}$ – множина зв'язків між вузлами множин $V_m$ та $V_k$, $m \neq k$, $m,k = \overline{1,M}$, $M$ – кількість шарів (взаємодіючих систем) БШМ. Множину

$$V^M = \bigcup_{m=1}^{M} V_m$$

називатимемо загальною сукупністю вузлів, а множину

$$E^M = (\bigcup_{m=1}^{M} E_m) \bigcup (\bigcup_{m,k=1, m \neq k}^{M} E_{mk})$$

загальною сукупністю ребер багатошарової мережі, $N^M$, $L^M$ – кількості елементів множин $V^M$ та $E^M$ відповідно.

Багатошарова мережа $G^M$ повністю описується матрицею суміжності

$$\mathbf{A}^M = \{\mathbf{A}^{km}\}_{m,k=1}^{M},$$

у якій значення $a_{ij}^{km} = 1$, якщо існує ребро, поєднуюче вузли $n_i^k$ та $n_j^m$, і $a_{ij}^{km} = 0$, $i, j = \overline{1, N^M}$, якщо такого ребра немає. При цьому блоки $\mathbf{A}^{mm}$ описують структуру внутрішньо шарових взаємодій у $m$-му шарі, а блоки $\mathbf{A}^{km}$ – структуру міжшарових взаємодій між $m$-тим та $k$-тим шарами БШМ, $m \neq k$, $m,k = \overline{1,M}$. Якщо всі блоки матриці $\mathbf{A}^M$ визначені для тотальної сукупності вузлів БШМ, то знімається проблема координації номерів вузлів у випадку їх незалежної нумерації у кожному шарі.

Більшість реально існуючих міжсистемних взаємодій є багатоцільовими та багатофункціональними. Це насамперед виражається у мультипотоковості таких утворень, тобто забезпеченні руху різних типів потоків. У ТСМ структура подібних міжсистемних взаємодій відображається так званими багатовимірними мережами [15]. Багатовимірна мережа є БШМ, у якій кожний шар відображає структуру системи, що забезпечує рух загалом відмінного від інших шарів типу потоку. Розглянемо у якості прикладу загальну транспортну систему, яка забезпечує рух двох основних видів потоків – пасажирських та вантажних, тобто її структуру можна зобразити у вигляді двовимірної мережі. Особливістю цієї структури, як і більшості багатовимірних мереж, є неможливість переходу потоку з одного шару на інший (перетворення пасажирів у вантажі і навпаки). Для спрощення аналізу процесу міжсистемних взаємодій у двовимірній загальній транспортній системі її можна поділити на дві чотиришарові монопотокові БШМС, шари якої (залізничний, автомобільний, авіаційний та водний) забезпечує рух лише одного типу потоку – пасажирського або вантажного. Характерною рисою монопотокових транспортних БШМС є відмінність носіїв потоків у кожному шарі (поїзди, автотранспортні засоби, літаки, кораблі). Загалом, під час деталізації структури реальних багатовимірних мереж спочатку доцільно виділяти шари, які забезпечують рух різних типів потоків, а потім кожний із таких монопотокових шарів зображати у вигляді БШМС, кожний шар якої забезпечує рух цих потоків специфічним носієм або системою-оператором. Спілкування у соціальних та інших інформаційно-комунікаційних мережах здійснюється шляхом обміну інформаційними потоками. Тобто, такі утворення можна вважати монопотоковими багатошаровими системами. Окремі шари таких систем зазвичай відображають процес функціонування різних систем-операторів інформаційних потоків, як це відбувається у системах мобільного та стаціонарного зв'язку, кабельного та супутникового телебачення, електронної та звичайної пошти, соціальних мереж тощо.

Потокову модель монопотокової БШМС зобразимо [16] у вигляді матриці суміжності $\mathbf{V}^M(t)$, елементи якої визначаються об'ємами потоків, які пройшли ребрами БШМ (1) за період $[t-T, t]$ до поточного моменту часу $t \geq T$:

$$\mathbf{V}^M(t) = \{V_{ij}^{km}(t)\}_{i,j=1}^{N}, {}_{k,m=1}^{M} \quad V_{ij}^{km}(t) = \frac{\widetilde{V}_{ij}^{km}(t)}{\max\limits_{s,g=\overline{1,M}} \max\limits_{l,p=\overline{1,N^M}} \{\widetilde{V}_{lp}^{sg}(t)\}}, \quad (2)$$

де

$$\widetilde{V}_{ij}^{km}(t) = \int\limits_{t-T}^{t} v_{ij}^{km}(\tau)d\tau; \quad v_{ij}^{km}(t) = \int\limits_{(n_i^k, n_j^m)} \rho_{ij}^{km}(t, \mathbf{x})dl; \quad \mathbf{\rho}(t,x) = \{\rho_{ij}^{km}(t, \mathbf{x})\}_{i,j=1}^{N^M}, {}_{k,m=1}^{M}$$

і $\rho_{ij}^{km}(t, \mathbf{x})$ – щільність потоку, який пересувається ребром $(n_i^k, n_j^m)$ БШМС у поточний момент часу $t > 0$, $\mathbf{x} \in (n_i^k, n_j^m) \subset R^n$, $n = 2, 3, ...$, $i, j = \overline{1, N^M}$, $k, m = \overline{1, M}$. Елементи потокової матриці суміжності БШМС визначаються на підставі емпіричних даних про рух потоків її ребрами. Натепер за допомогою сучасних засобів відбору інформації такі дані достатньо легко отримати для багатьох природних та переважної більшості створених людиною систем, у тому числі інформаційних [17, 18, 19]. Матриця $\mathbf{V}^M(t)$ має блочну структуру, у якій діагональні блоки $\mathbf{V}^{mm}(t)$ описують об'єми руху внутрішньо шарових потоків у $m$-му шарі, а позадіагональні блоки $\mathbf{V}^{km}(t)$ – об'єми руху потоків між $m$-тим та $k$-тим шарами БШМС, $m \neq k$, $m, k = \overline{1, M}$.

Ми обчислюємо значення елементів матриці $\mathbf{V}^M(t)$ на часовому інтервалі $[t-T, t]$, $t \geq T$, для того, щоб нівелювати випадкові збурення, які можуть виникнути в процесі руху потоків в окремі моменти часу. Ці значення є динамічними величинами, оскільки визначаються до поточного моменту часу, а отже, неперервно змінюються. Тривалість проміжку $T$ залежить від динаміки поведінки системи. Наприклад, для перерахованих у попередній секції масових соціальних збурень, які відбувалися у багатьох країнах світу, цей проміжок зазвичай не повинен перевищувати однієї доби. Свідченням цього є те, що навіть 29 листопада 2013 року практично ніхто не міг передбачити, що побиття студентів у Києві у ніч на 30 листопада практично на наступний день спровокує появу нового Майдану в Україні. Для моделей поширення Covid-19 проміжок $T$ (для згладжування різниці у кількості нововиявлених інфікованих у будні та вихідні дні) зазвичай дорівнював тижню [20].

Спільноти можуть виникати, як в окремих шарах-системах, так і у БШМС загалом. При цьому вони можуть як посилювати, накладаючись або перетинаючись, так і нівелювати одна одну. Так, письменницьке співтовариство достатньо чітко поділяється на спільноти авторів детективних, фантастичних, історичних та інших за жанрами творів. Однак, автор кримінальних романів може привносити у них елементи фантастики або історичних подій, фантаст – детективну або історичну складову тощо. Тобто, спільноти різних за жанрами письменників часто перетинаються. Водночас, спільноти спортивного співтовариства, які складаються із спортсменів різних видів спорту, перетинаються значно менше. Дійсно, рідко коли професійний футболіст на такому ж рівні займається важкою атлетикою, а важкоатлет – шахами чи бігом на марафонську дистанцію. Це означає доцільність розпочинати пошук спільнот саме з окремих шарів багатошарової мережевої системи.

## 4. Спільноти у шарах багатошарових мережевих систем

Для спрощення викладу позначимо у цій секції потокову матрицю суміжності довільного шару-системи БШМС через $\mathbf{V}(t) = \{V_{ij}(t)\}_{i,j=1}^{N}$, елементи якої $V_{ij}(t)$ дорівнюють відносним об'ємам потоків, які пройшли ребром $(n_i, n_j)$, $i, j = \overline{1, N}$, цього шару за проміжок часу $[t-T, t]$, $t \geq T$, $N$ – кількість вузлів шару. Розглянемо два підходи до визначення спільнот у такій мережевій системі.

### 4.1. Пошук спільнот на підставі ієрархії вкладеності

У реальних великих системах першими «кандидатами» в спільноти є підсистеми різних рівнів ієрархії, побудованої за принципом вкладеності, коли менше входить до складу більшого [21] (рис. 1). Дійсно, учні класу більше спілкуються між собою, ніж з учнями інших класів чи курсів, люди певних професій – частіше, ніж з представниками інших спеціальностей, представники різних соціальних груп або вікових категорій також віддають перевагу спілкуванню із людьми тих же груп або категорій. Тобто, виділені за певною ознакою однорідності елементів, підсистеми МС мають більшу ймовірність для утворення спільноти.

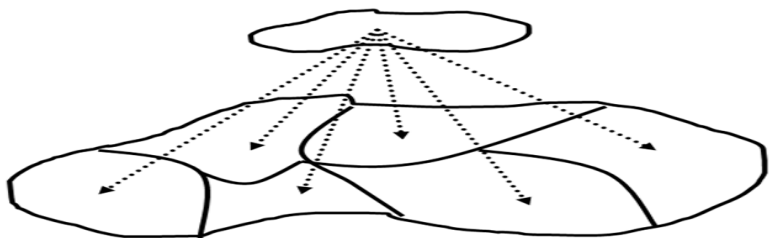

**Рис. 1.** Приклад ієрархії, побудованої за принципом вкладення

Нехай вихідна мережева система $S$ поділяється за принципом вкладення на $L$ підсистем найнижчого рівня ієрархії

$$S_l \subset S = \bigcup_{l=1}^{L} S_l,$$

множини вузлів $H_{S_l} = \{n_i^l\}_{i=1}^{N_l}$ яких не перетинаються, $N_l$ – кількість вузлів підсистеми $S_l$, $l = \overline{1, L}$. Позначимо $G_{S_l}^{out}$ – сукупність усіх вузлів-генераторів потоків, що поширюються мережевою системою $S$, які входять до множини $H_{S_l}$. Визначимо за допомогою параметра

$$\xi_{S_l}^{out}(t) = \sum_{i \in G_{S_l}^{out}} \xi_i^{out}(t) / s(\mathbf{V}(t))$$

силу впливу підсистеми $S_l$ на МС загалом. Тут $\xi_i^{out}(t)$ – об'єм вихідних потоків, що генеруються у вузлі $n_i$ із множини $G_{S_l}^{out}$ та поширюються мережевою системою $S$ і $s(\mathbf{V}(t))$ – сумарний об'єм потоків, які пройшли системою $S$ за проміжок часу $[t–T, t]$, тобто

$$s(\mathbf{V}(t)) = \sum_{i,j=1}^{N} V_{ij}(t), \ t \geq T.$$

Нехай

$$R_{S_l}^{out} = \bigcup_{i \in G_{S_l}^{out}} R_{l,i}^{out}$$

є множиною номерів вузлів – кінцевих приймачів (не транзитерів), згенерованих у вузлах множини $G_{S_l}^{out}$ потоків. Розділимо множину $R_{S_l}^{out}$ на дві підмножини, а саме

$$R_{S_l}^{out} = R_{S_l, int}^{out} \bigcup R_{S_l, ext}^{out},$$

де $R_{S_l, int}^{out}$ – підмножина вузлів $R_{S_l}^{out}$, які належать $H_{S_l}$, та $R_{S_l, ext}^{out}$ – підмножина вузлів $R_{S_l}^{out}$, які належать доповненню до $H_{S_l}$ у вихідній мережевій системі *S*. Множину $R_{S_l, ext}^{out}$ називатимемо областю вихідного впливу підсистеми $S_l$ на МС. Зовнішню та внутрішню вихідну силу впливу вузлів-генераторів потоків, що належать множині $G_{S_l}^{out}$, на підмережі $R_{S_l, ext}^{out}$ та $R_{S_l, int}^{out}$ визначимо за допомогою параметрів

$$\xi_{S_l, ext}^{out}(t) = \sum_{i \in R_{S_l, ext}^{out}} \xi_i^{out}(t) / s(\mathbf{V}(t))$$

та

$$\xi_{S_l, int}^{out}(t) = \sum_{i \in R_{S_l, int}^{out}} \xi_i^{out}(t) / s(\mathbf{V}(t))$$

відповідно. Тоді величина

$$\varpi_{S_l}^{out}(t) = \frac{\xi_{S_l, ext}^{out}(t)}{\xi_{S_l, int}^{out}(t)}$$

визначає відносну силу впливу підсистеми $S_l$ на мережеву систему загалом. А саме, чим меншим є значення параметра $\varpi_{S_l}^{out}(t)$, $t \geq T$, тим меншим є сила впливу підсистеми $S_l$ на МС. Іншими словами, у такому випадку потокові зв'язки між вузлами-генераторами та вузлами – кінцевими приймачами потоків є значно сильнішими у межах підсистеми $S_l$, $l = \overline{1, L}$, ніж між ними та іншими вузлами – приймачами потоків системи *S* загалом.

Позначимо $G_{S_l}^{in}$ – сукупність усіх вузлів – кінцевих приймачів потоків, які входять до множини $H_{S_l}$. Визначимо за допомогою параметра

$$\xi_{S_l}^{in}(t) = \sum_{i \in G_{S_l}^{in}} \xi_i^{in}(t) / s(\mathbf{V}(t))$$

силу впливу мережевої системи *S* на підсистему $S_l$. Тут $\xi_i^{in}(t)$ – об'єм вхідних потоків, які генеруються вузлами системи *S* та кінцево приймаються у вузлі $n_i$ із множини $G_{S_l}^{in}$. Нехай

$$R_{S_l}^{in} = \bigcup_{i \in G_{S_l}^{in}} R_{l,i}^{in}$$

є множиною номерів вузлів-генераторів, потоки з яких кінцево приймаються у вузлах множини $G_{S_l}^{in}$. Розділимо множину $R_{S_l}^{in}$ на дві підмножини, а саме

$$R_{S_l}^{in} = R_{S_l, int}^{in} \bigcup R_{S_l, ext}^{in},$$

де $R_{S_l, int}^{in}$ – підмножина вузлів $R_{S_l}^{in}$, які належать $H_{S_l}$, та $R_{S_l, ext}^{in}$ – підмножина вузлів $R_{S_l}^{in}$, які належать доповненню до $H_{S_l}$ у вихідній мережевій системі *S*. Множину $R_{S_l, ext}^{in}$ називатимемо областю вхідного впливу мережевої системи *S* на підсистему *S_l*. Зовнішню та внутрішню вхідну силу впливу вузлів – кінцевих приймачів потоків, що належать множині $G_{S_l}^{in}$, на підмережі $R_{S_l,ext}^{in}$ та $R_{S_l,int}^{in}$ визначимо за допомогою параметрів

$$\xi_{S_l,ext}^{in}(t) = \sum_{i \in R_{S_l,ext}^{in}} \xi_i^{in}(t) / s(\mathbf{V}(t))$$

та

$$\xi_{S_l,int}^{in}(t) = \sum_{i \in R_{S_l,int}^{in}} \xi_i^{in}(t) / s(\mathbf{V}(t))$$

відповідно. Тоді величина

$$\varpi_{S_l}^{in}(t) = \frac{\xi_{S_l,ext}^{in}(t)}{\xi_{S_l,int}^{\text{int}}(t)}$$

визначає відносну силу впливу мережевої системи *S* на підсистему *S_l*. А саме, чим меншим є значення параметра $\varpi_{S_l}^{in}(t)$, $t \geq T$, тим меншим є сила впливу системи *S* на підсистему *S_l*. Іншими словами, у такому випадку потокові зв'язки між вузлами – кінцевими приймачами та вузлами-генераторами потоків є значно сильнішими у межах підсистеми *S_l*, $l = \overline{1,L}$, ніж між ними та іншими вузлами-генераторами потоків системи *S* загалом.

Природно вважати, що чим більші об'єми руху потоків між двома вузлами МС, тим сильніший взаємозв'язок між ними. Це твердження визначає достатньо обґрунтований критерій існування спільноти в межах певної групи вузлів (підсистеми) МС. Тому пара параметрів $(\varpi_{S_l}^{out}(t), \varpi_{S_l}^{in}(t))$, а саме сумісне виконання умов

$$\varpi_{S_l}^{out}(t) \leq \varpi_c \ll 1, \quad \varpi_{S_l}^{in}(t) \leq \varpi_c \ll 1, \qquad (3)$$

де $\varpi_c$ – наперед задана величина, дає змогу визначити достатньо об'єктивний критерій того, що підсистема *S_l* утворює спільноту у мережевій системі *S*. Дійсно, чим меншим є значення цих параметрів, тим меншою є зовнішня взаємодія підсистеми *S_l*, $l = \overline{1,L}$, із системою загалом та більшими взаємодії усередині підсистеми, що і є, по-суті, визначенням спільноти.

Узагальнюючи, оскільки за визначенням спільнотою вважається деяка група вузлів (підсистема *S\**) системи *S*, зв'язки між якими є щільнішими (структурний показник) та сильнішими (функціональний показник), ніж між ними та іншими вузлами мережевої системи *S*, то об'єктивними критеріями існування спільноти в межах підсистеми *S\** можна вважати:
1) показник відносної щільності зв'язків (ребер) підсистеми *S\**, який обчислюється за формулою

$$\mu(S^*, S) = \left(\frac{L^*}{N^*}\right) \bigg/ \left(\frac{L}{N}\right)$$

де *N\** – кількість вузлів та *L\** – кількість зв'язків підсистеми *S\**, *N* – кількість вузлів та *L* – кількість зв'язків системи *S*;
2) показник відносної потокової сили взаємозв'язків підсистеми *S\**, який обчислюється за формулою

$$\theta(S^*, S) = \left(\frac{s(\mathbf{V}_{S^*}(t))}{N^*}\right) \bigg/ \left(\frac{s(\mathbf{V}(t))}{N}\right),$$

та визначає відношення питомої сили взаємозв'язків між вузлами у підсистемі $S^*$ та системі $S$ відповідно. Якщо показники питомої щільності та потокової сили взаємозв'язків для підсистема $S^*$ значно перевищують відповідні показники для системи $S$, то можна цілком обґрунтовано вважати, що підсистема $S^*$ утворює спільноту у вихідній мережевій системі. Очевидно, що поняття «значно перевищують» є достатньо розмитим. Тому доцільно визначати конкретні значення $\mu(S^*, S)$ та $\theta(S^*, S)$, які встановлюють рівень такого перевищення, а саме, виконання умов

$$\mu(S^*, S) \geq \mu^* \gg 1 \qquad (4)$$

та/або

$$\theta(S^*, S) \geq \theta^* \gg 1, \qquad (5)$$

де $\mu^*$ та $\theta^*$ – наперед визначені величини. Зазначимо, що умова (5) є слабшою, ніж умова (3) та загалом не накладає таких жорстких обмежень на взаємодію підсистеми $S^*$ з іншою частиною системи $S$. Окрім того, значення

$$\nu(S^*, S) = \frac{s(\mathbf{V}_{S^*}(t))}{s(\mathbf{V}(t))}$$

дає змогу відстежувати роль підсистеми $S^*$ у процесі функціонування мережевої системи $S$ та динаміку зміни цієї ролі у часі. У випадку, якщо на даному рівні ієрархії, побудованої за принципом вкладення, спільнот не виявлено, то переходимо до наступного, вищого рівня цієї ієрархії.

Основним недоліком розглянутого вище методу є те, що він орієнтований на виявлення спільнот, сформованих протягом достатньо тривалого проміжку часу, який дозволяє побудувати відповідні ієрархії вкладення. Розглянемо метод, який дозволяє відстежувати появу та динамічний розвиток спільноти у мережевій системі, орієнтуючись насамперед на переважаючу силу взаємозв'язків між її елементами.

### 4.2. Спільноти та потокові серцевини мережевих систем

Введемо поняття потокової $\lambda$-серцевини мережевої системи [22], як найбільшої підсистеми вихідної МС, для якої елементи потокової матриці суміжності $\mathbf{V}(t)$ задовольняють нерівності

$$V_{ij}(t) \geq \lambda, \ i, j = \overline{1, N}, \ t \geq T, \lambda \in [0,1].$$

Поняття потокової серцевини МС дозволяє побудувати на підставі критерію (5) наступний алгоритм виявлення спільнот у мережевій системі (рис. 2а – структура вихідної МС, 2б – вихідна МС з відображеними об'ємами руху потоків протягом періоду тривалістю $T$, товщина ліній є пропорційною об'ємам потоків):
1) приймаємо значення $i = 0$, $S = S_{\lambda_0}$, $\lambda_0$ – мінімальне відмінне від 0 значення параметра $\lambda \in [0,1]$;
2) поступово збільшуємо значення $\lambda$ поки для певного $\lambda = \lambda_{i+1}$ не буде виконана умова (3) або (5) чи принаймні одна із виділених раніше складових $\lambda_i$-серцевини не поділиться на незв'язні складові (рис. 2в – 2г);
3) якщо значення $\lambda_{i+1} < 1$, то приймаємо $i = i + 1$ та переходимо до пункту 2, інакше закінчуємо виконання алгоритму.

Регулюючи значення $T$ у сторону зменшення або збільшення, ми робимо процедуру пошуку спільнот за допомогою потокової моделі МС та методу $\lambda$-серцевин більш або менш чутливою до швидкоплинних змін у структурі виявлених спільнот.

Зазначимо, що для зображеної на рис. 2а регулярної мережі критерій (4) не виконується для жодної її підмережі. Однак, у випадку нерегулярних мереж цей критерій можна використовувати для перевірки, чи справді зв'язки у виділеній за допомогою критерію (5) підсистемі МС справді є щільнішими, ніж у середньому по мережі. Тобто, функціональний критерій (3) або (5) дають змогу виділяти спільноти у мережевій системі для яких відомі структурні методи не працюють. Очевидно, що структура та склад вузлів і зв'язків виділених за допомогою описаного вище алгоритму спільнот легко визначається із матриці $\mathbf{V}(t)$, $t \geq T$.

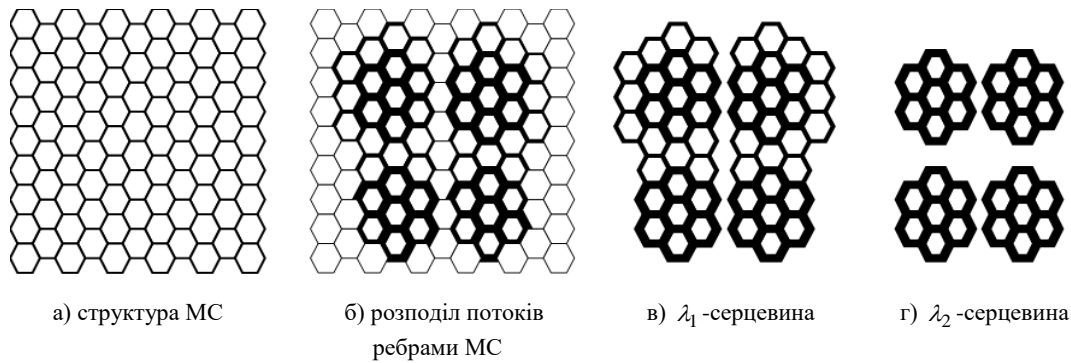

а) структура МС    б) розподіл потоків ребрами МС    в) $\lambda_1$-серцевина    г) $\lambda_2$-серцевина

**Рис. 2.** Застосування потокових $\lambda$-серцевин для виділення спільнот у мережі

Очевидно, що потокова модель МС дає змогу здійснювати неперервний моніторинг зміни обсягів потоків, які пересуваються між вузлами мережевої системи. Це дозволяє практично у реальному часі відстежувати процеси появи та розвитку спільнот у мережі, що набагато важче здійснити за допомогою структурних методів дослідження. Так, Донбас до 2014 р. був одним із найбільш промислово розвинутих регіонів України з дуже тісними зв'язками між шахтами, родовищами, гірничо-збагачувальними та металургійними підприємствами, які знаходились на його території. Це супроводжувалось необхідністю існування набагато щільнішої, ніж у середньому по країні, транспортної та енергетичної інфраструктури. Загалом таке утворення можна вважати промисловою спільнотою. Однак, після 2014 р. унаслідок закриття значної частини шахт та родовищ і припинення роботи багатьох металургійних комбінатів ця спільнота практично перестала існувати, хоча щільна транспортна та енергетична мережі нікуди не зникли. Можна навести чимало подібних прикладів: автопромисловий центр у Детройті, вугільна промисловість у Великобританії та Германії, виноробство у Франції та багато інших промислових регіонів, попит на продукцію, яка вироблялась у них, поступово зменшився та зник або були вичерпані поклади корисних копалин, які видобувались у цих регіонах. Зі структурного погляду згідно критерію (4) підсистема може утворювати спільноту, але з функціонального погляду згідно критерію (3) або (5) вона її не утворює, і навпаки.

Зазначимо, що тут і нижче ми навмисно використовуємо такі прості приклади структур мережевих систем, оскільки саме на них відомі числові та візуальні структурні методи пошуку спільнот практично не спрацьовують. Подібні приклади можна навести і для значно складніших реальних мережевих структур, наприклад, системи взаємозв'язків між регіонами України (рис. 3), які також загалом є регулярними, незважаючи на візуальну складність.

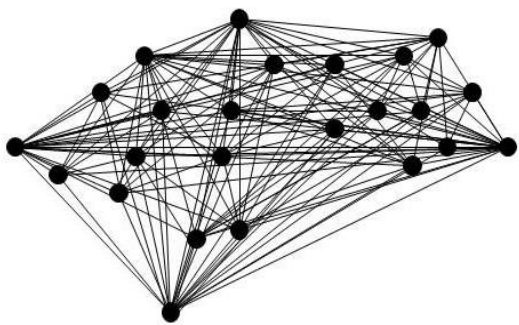

**Рис. 3.** Структура економічних взаємозв'язків між регіонами України

## 5. Агрегат-мережі та серцевини багатошарових мережевих систем

Локальною характеристикою $\varepsilon_{ij}$ ребра $(n_i, n_j)$ загальної сукупності ребер $E^M$ багатошарової мережі, де $n_i$ та $n_j$ – вузли із загальної сукупності вузлів $V^M$, яку називатимемо його структурною агрегат-вагою, є кількість шарів, у яких таке ребро є присутнім. Структурною агрегат-вагою $\varepsilon_{ii}$ вузла $n_i$ БШМ є кількість шарів, до складу яких він входить, $i, j = \overline{1, N^M}$. Для довільної багатошарової мережі матриця суміжності $\mathbf{E} = \{\varepsilon_{ij}\}_{i,j=1}^{N^M}$ повністю визначає зважену мережу, яку називатимемо структурною агрегат-мережею БШМ. Елементи матриці $\mathbf{E}$ визначають інтегральні структурні характеристики вузлів та ребер багатошарової мережі (рис. 4). Для монопотокових БШМ вага кожного ребра відображає кількість можливих носіїв або систем-операторів, які можуть забезпечити рух відповідного типу потоку, а вага кожного вузла – кількість систем, до складу яких він входить.

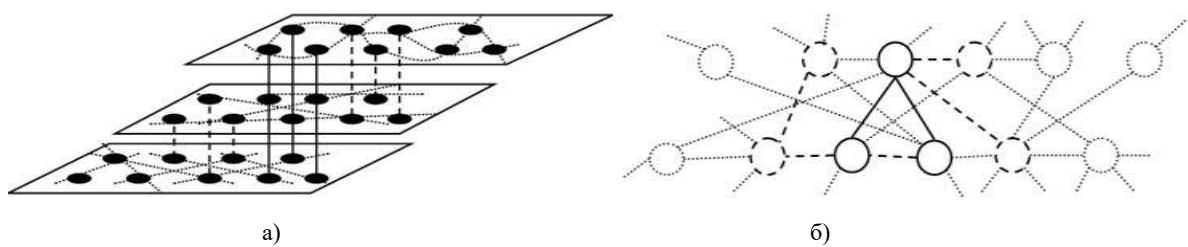

**Рис. 4.** Фрагмент тришарової БШМ (а) та її агрегат-мережі (б – ──── – для $\varepsilon_{ij} = 3$, ─ ─ ─ – для $\varepsilon_{ij} = 2$, ….. – для $\varepsilon_{ij} = 1$, $i, j = \overline{1, N^M}$)

Перехід до агрегат-мереж можна використати для розроблення структурних методів пошуку спільнот у БШМ [7]. Структурною $p_{ag}$-серцевиною агрегат-мережі БШМС називатимемо мережу, елементи матриці суміжності якої визначаються за співвідношенням

$$\varepsilon_{ij}^{p_{ag}} = \begin{cases} \varepsilon_{ij}(t), \text{ якщо } \varepsilon_{ij} \geq p_{ag}, \\ 0, \quad \text{ якщо } \varepsilon_{ij} < p_{ag}, \ i, j = \overline{1, N^M}. \end{cases}$$

Зі зростанням значення $p_{ag}$ можна вважати, що $p_{ag}$-серцевини є одними з найбільш ймовірних «кандидатів» у спільноти, оскільки дублювання зв'язків у МС зазвичай відбувається за двох причин, а саме, коли ці зв'язки є достатньо важливими для системи та якщо через них відбувається розподіл переміщення достатньо великих об'ємів потоків.

Потоковою агрегат-мережею БШМС, називатимемо мережеву систему, елементи матриці суміжності $\mathbf{F}(t) = \{f_{ij}(t)\}_{i,j=1}^{N^M}$ якої визначаються за співвідношеннями

$$f_{ij}(t) = \sum_{k,m=1}^{M} V_{ij}^{km}(t) / M^2, \ i, j = \overline{1, N^M}, \ t \geq T.$$

Елементи матриці $\mathbf{F}(t)$ визначають інтегральні потокові характеристики вузлів та ребер багатошарової мережевої системи. Потокову $\lambda_{ag}$-серцевину агрегат-мережі монопотокової БШМС визначимо за допомогою матриці суміжності, елементи якої обчислюємо за співвідношенням

$$f_{ij}^{\lambda_{ag}}(t) = \begin{cases} f_{ij}(t), \text{ якщо } f_{ij}(t) \geq \lambda_{ag}, \\ 0, \quad \text{ якщо } f_{ij}(t) < \lambda_{ag}, \ i, j = \overline{1, N^M}, t \geq T, \ \lambda_{ag} \in [0, 1]. \end{cases}$$

Для пошуку спільнот у агрегат-мережі БШМС можна використовувати алгоритм, описаний у секції 4.2. Так, на рис. 5а – 5в зображені спільноти, які містяться у 1-3 шарах тришарової БШМС відповідно. На рис. 5г відображена потокова агрегат-мережа БШМС, виділена в момент часу $t \geq T$. Рис. 5д та 5е містять зображення спільнот, отриманих за допомогою методу $\lambda$-серцевин для $\lambda_{ag} = \lambda_1$ та $\lambda_{ag} = \lambda_2$.

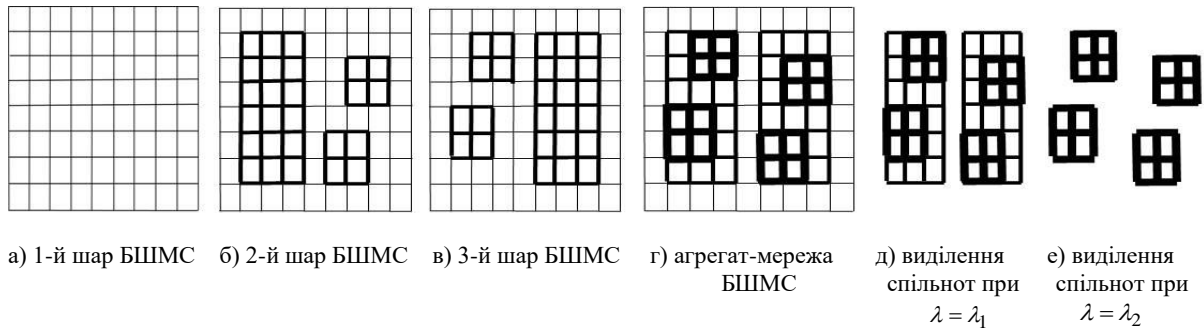

а) 1-й шар БШМС    б) 2-й шар БШМС    в) 3-й шар БШМС    г) агрегат-мережа БШМС    д) виділення спільнот при $\lambda = \lambda_1$    е) виділення спільнот при $\lambda = \lambda_2$

**Рис. 5.** Застосування методу потокових $\lambda$-серцевин для виділення спільнот у агрегат-мережі БШМС

Застосування методу потокових $\lambda$-серцевин для агрегат-мережі БШМС дає змогу визначати наявність спільнот у багатошаровій системі, насамперед тих, які одночасно формуються у різних шарах та у результаті процесу міжсистемних взаємодій. Однак, визначити конкретний внесок кожного із шарів у створенні таких спільнот та інтенсивність взаємодій між ними, як складових різних систем, як і структурні методи пошуку спільнот, які базуються на використанні поняття її $p_{ag}$-серцевини, цей метод не може. Приклад такої ситуації для тришарової мережевої системи відображений на рис. 6. Зокрема, одна із спільнот є повністю сформованою у першому шарі та практично не існує у інших шарах (рис. 6а, лівий нижній кут). Друга із спільнот сформована у всіх шарах БШМС (рис. 6а – 6в, правий верхній кут), але виділяється у них порівняно слабо. Однак, ця спільнота є співмірною за силою взаємозв'язків із першою у агрегат-мережі багатошарової системи загалом (рис. 6г). До подібних спільнот умовно можна віднести згадані вище приклади письменницького та наукового співтовариств.

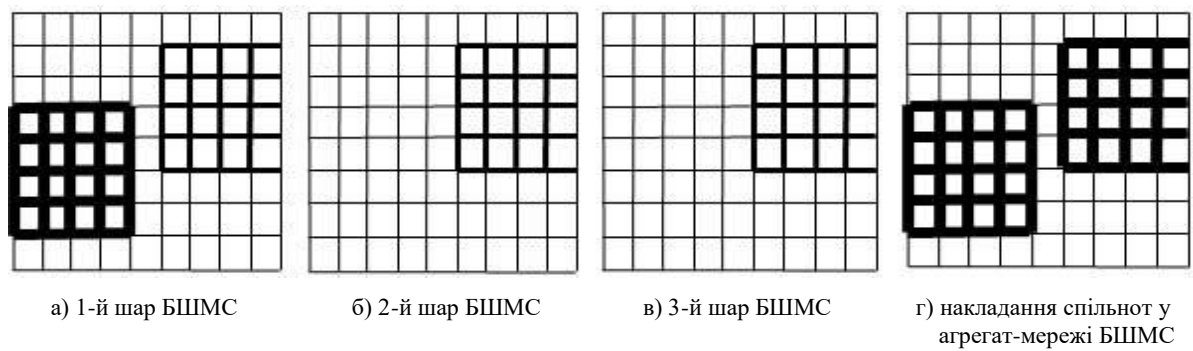

а) 1-й шар БШМС    б) 2-й шар БШМС    в) 3-й шар БШМС    г) накладання спільнот у агрегат-мережі БШМС

**Рис. 6.** Спільноти в окремих шарах БШМС та її потоковій агрегат-мережі

Іншим недоліком методу потокових $\lambda$-серцевин для агрегат-мережі БШМС є можливість нівелювання спільнот, які існують в окремих шарах (рис. 7), що не сприяє кращому розумінню процесів, які перебігають у багатошаровій системі. Прикладом такої ситуації є вже згадане у секції 3 спортивне співтовариство, у якому окремі спільноти за видами спорту практично не перетинаються. У цьому випадку незалежні спільноти, які існують у 1-3 шарах тришарової БШМС (рис. 7а – 7в) практично «зливаються» у її агрегат-мережі, утворюючи єдину спільноту (рис. 7г), що насправді не відповідає дійсності. Зазначимо, що визначити конкретний внесок кожного із шарів у «злитті» таких спільнот, як і структурні методи, які базуються на використанні поняття її $p_{ag}$-серцевини, цей метод не може. До того ж, потокові $\lambda_{ag}$-серцевини агрегат-мережі БШМС не дають змогу встановлювати міжсистемні взаємодії між спільнотами, які існують у різних шарах багатошарової системи. Тому розроблення методів пошуку спільнот у БШМС, зокрема виділення таких утворень в окремих шарах та встановлення взаємодій між ними є не менш важливим. Очевидно, що такі спільноти зазвичай також мають вигляд багатошарової системи.

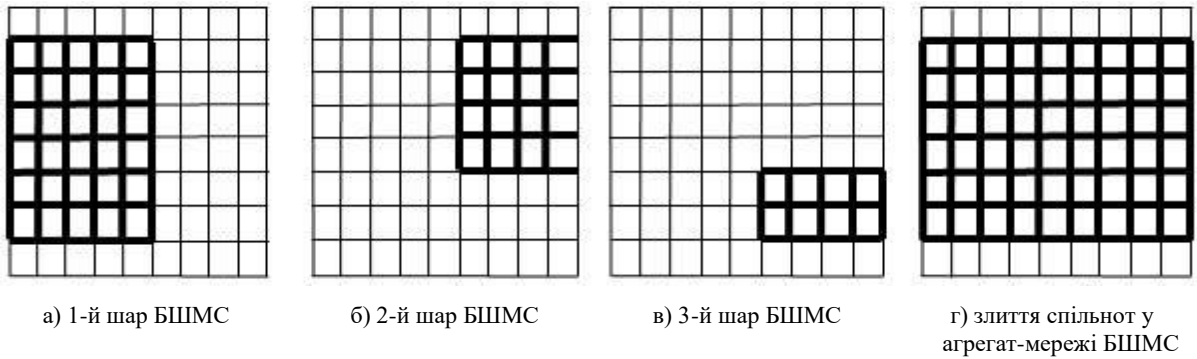

а) 1-й шар БШМС    б) 2-й шар БШМС    в) 3-й шар БШМС    г) злиття спільнот у агрегат-мережі БШМС

**Рис. 7.** Нівелювання спільнот, які існують в окремих шарах, в агрегат-мережі багатошарової мережевої системи

## 6. Виділення спільнот у багатошарових мережевих системах

Визначимо поняття потокової $\lambda$-серцевин багатошарової мережевої системи. Сформуємо матрицю суміжності $V_\lambda^M(t) = \{V_{\lambda,ij}^{mk}(t)\}_{i,j=1\ m,k=1}^{N^N\ M}$ у якій

$$V_{\lambda,ij}^{mk}(t) = \begin{cases} V_{ij}^{mk}(t), & \text{якщо } V_{ij}^{mk}(t) \geq \lambda, \\ 0, & \text{якщо } V_{ij}^{mk}(t) < \lambda, \ i,j = \overline{1,N^M}, \ m,k = \overline{1,M}, \ t \geq T, \lambda \in [0,1]. \end{cases}$$

Аналогічно секції 4.2 будується алгоритм визначення спільнот у БШМС у якому послідовно зі збільшенням значення $\lambda$ виділяються спільноти в окремих шарах багатошарової системи та зв'язки між ними. На рис. 8а відображено фрагмент тришарової мережевої системи, а на рис. 8б – виділена за допомогою цього алгоритму її потокова $\lambda_1$-серцевина, потокові зв'язки між вузлами якої за критерієм (5) є принаймні у три рази сильнішими, ніж у середньому по БШМС. Очевидно, що агрегат-мережа потокової $\lambda$-серцевини багатошарової системи входить до складу її $\lambda_{ag} = \lambda$-серцевини, тобто спільноти, які виділяються методом $\lambda$-серцевин для БШМС загалом зазвичай є підспільнотами спільнот, які отримуються методом $\lambda$-серцевин для агрегат-мережі БШМС.

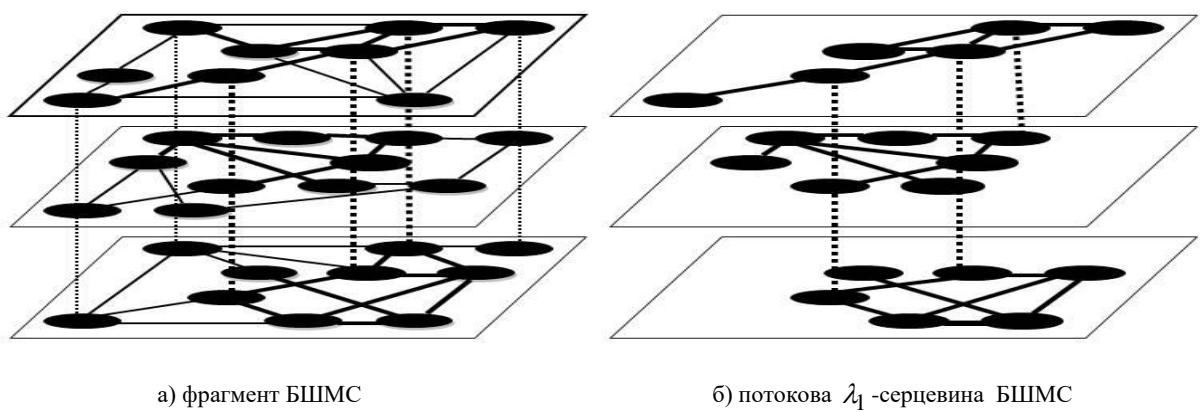

а) фрагмент БШМС    б) потокова $\lambda_1$-серцевина БШМС

**Рис. 8.** Виділення спільнот у БШМС загалом методом $\lambda$-серцевин

Проектуючи отриману за допомогою описаного вище методу $\lambda$-серцевину БШМС на її агрегат-мережу, ми можемо визначати об'єднання, перекривання або нівелювання спільнот, які існують в окремих її шарах.

## 7. Висновки

Вивчення явищ зародження та розвитку спільнот сприяє кращому розумінню процесів функціонування реальних складних мережевих систем та міжсистемних взаємодій, які існують у фізичному світі, живій природі та людському соціумі. Саме тому в останні десятиліття цій проблематиці присвячено чимало наукових досліджень. Структурний підхід до виявлення спільнот у складних мережевих та багатошарових мережевих системах, який натепер розвивається у межах теорії складних мереж, має низку недоліків, серед яких найпершим слід назвати відсутність обґрунтованого критерію, що зв'язки у межах виявленого утворення, яке вважається спільнотою, є не тільки щільнішими, але й сильнішими, ніж у середньому по мережі. На противагу структурному, потоковий підхід дає змогу ефективно вирішувати цю проблему, адже твердження, що чим більші обсяги потоків поєднують два вузли мережі, тим сильніший зв'язок між ними, видається цілком обґрунтованим. Динамічність утворення та розвитку спільнот у сучасному людському суспільстві, принаймні деяких з яких несуть явну або приховану загрозу суспільному спокою та безпеці, робить задачу своєчасного виявлення таких утворень ще більш актуальною. Запропоновані у статті методи виявлення спільнот, які базуються на використанні понять потокової серцевини мережевих та агрегат-мереж і потокових серцевин багатошарових мережевих систем, робить цю проблему вирішуваною навіть у режимі реального часу. Додатковою перевагою пропонованих методів є можливість їх застосування у тих випадках, коли структура мережевої або багатошарової мережевої системи робить інші відомі підходи практично непрацездатними.

## 8. Перелік посилань